\documentclass[10pt]{article}
\usepackage{graphicx}
\usepackage{cite}    
\begin{document}

\newcommand{\be}{\begin{equation}}
\newcommand{\ee}{\end{equation}}
\newcommand{\bea}{\begin{eqnarray}}
\newcommand{\eea}{\end{eqnarray}}
\newcommand{\bi}{\bibitem}
\newcommand{\la}{\langle}
\newcommand{\ra}{\rangle}
\renewcommand{\r}{({\bf r})}
\newcommand{\rp}{({\bf r'})}
\newcommand{\rpp}{({\bf r''})}
\newcommand{\rrp}{({\bf r},{\bf r}')}
\newcommand{\ua}{\uparrow}
\newcommand{\da}{\downarrow}
\newcommand{\s}{\sigma}
\newcommand{\eps}{\epsilon}

\title{Orbital-polarization terms: from a phenomenological to a 
first-principles description of orbital magnetism in density-functional theory}
\author{J. M. Morbec and K. Capelle\\  \\
Departamento de F\'{\i}sica e Inform\'atica\\
Instituto de F\'{\i}sica de S\~ao Carlos\\
Universidade de S\~ao Paulo\\
Caixa Postal 369, 13560-970 S\~ao Carlos, SP\\
Brazil}
\date{\today}

\maketitle

\begin{abstract}
Phenomenological orbital-polarization (OP) terms have been repeatedly 
introduced in the single-particle equations of spin-density-functional theory, 
in order to improve the description of orbital magnetic moments in systems 
containing transition metal ions. Here we show that these {\em 
ad hoc} corrections can be interpreted as approximations to the 
exchange-correlation vector potential ${\bf A}_{xc}$ of 
current-density-functional theory (CDFT). 
This connection provides additional information on both approaches: 
Phenomenological OP terms are connected to first-principles theory,
leading to a rationale for their empirical success and a reassessment 
of their limitations and the approximations made in their derivation. 
Conversely, the connection of OP terms with CDFT leads 
to a set of simple approximations to the CDFT potential ${\bf A}_{xc}$,
with a number of desirable features that are absent
from electron-gas-based functionals.
\end{abstract}

\newpage
\tableofcontents
\newpage

\section{\label{intro}Introduction}

Magnetic phenomena arising as a consequence of the spin degrees of freedom and
the antisymmetrization of the wave function are ubiquitous in quantum
chemistry and physics, and are routinely treated within both density-based 
and wave-function-based approaches. Magnetic phenomena arising from orbital
currents, on the other hand, are not automatically accounted for by standard
methods for electronic-structure calculations. The present paper is dedicated
to exploring links between a phenomenological and a first-principles
approach to orbital magnetism in density-functional theory (DFT). These
links shed new light on both approaches, and pave the way for new applications
of each.

In Sec.~\ref{orbmag}, of the present paper, we briefly recall a number of
distinct ways in which orbital magnetism can occur in a many-electron system.
In Sec.~\ref{treatments}, we describe two computational approaches to orbital 
magnetism. Section~\ref{cdft} is devoted to current-density functional theory 
(CDFT), as a first-principles method for the description of orbital magnetism 
in DFT \cite{vr1,vr2}. Independently of CDFT, a variety of 
phenomenological {\em orbital polarization} (OP) terms were proposed as 
{\em add-ons} 
to the single-particle equations of spin-density-functional theory (SDFT) 
\cite{brookspaper,eriksson,norman1,norman3,shick}. These terms are quite 
commonly used in band-structure calculations of magnetic solids, but are
much less known in quantum chemistry. We devote Sec.~\ref{brooks} to a quick 
description of the motivation and form of several such {\em add-on} terms.

In Sec.~\ref{connections}, we show that these phenomenological terms can be 
related to the first-principles approach, both for single-particle energies 
(Sec.~\ref{eps}) and total energies (Sec.~\ref{Es}). This connection benefits 
both sides: Phenomenological OP terms are connected to first-principles 
theory, leading to a rationale for their empirical success and a reassessment
of their limitations and the approximations made in their derivation.
Conversely, the connection of orbital-polarization terms with CDFT leads
to a set of simple approximations to the CDFT vector potential ${\bf A}_{xc}$ of
systems containing open-shell atoms, with a number of desirable features 
that are absent from electron-gas-based functionals.

Section~\ref{conclusions} contains our conclusions.

\section{\label{orbmag}Orbital magnetism in many-electron systems}

There are at least four conceptually distinct ways in which orbital
magnetism can appear in a physical system. One is the presence of external
magnetic fields ${\bf B}\r$, whose vector potential ${\bf A}\r$ enters the
Hamiltonian via the usual minimal substitution in the kinetic energy
\be
\frac{\hat{p}^2}{2m} \longrightarrow \frac{1}{2m}
\left(\hat{p}-\frac{q}{c}{\bf A}\r\right)^2.
\ee

A second way in which orbital magnetism can appear is due to
current-current interactions, which are part of the Breit interaction
\cite{strange,pyykko} and therefore a relativistic effect. 
The nonretarded part of this interaction is
\be
-\frac{q^2}{c^2} \int d^3 r \int d^3 r'\,
\frac{{\bf j}_p\r \cdot {\bf j}_p({\bf r'})}{|{\bf r}-{\bf r'}|}
=-\frac{q}{c} \int d^3 r \, {\bf j}_p\r \cdot {\bf A}_H\r,
\ee
which describes the Hartree-like coupling of currents to the self-induced
vector potential, corresponding to the Amperian currents of classical
electrodynamics. 
                                                                                
Third, in the presence of spin-orbit coupling a nonzero spin magnetization
can induce an orbital magnetization. Orbital magnetic moments induced by
spin-orbit coupling can be treated within relativistic DFT, and are often
described within SDFT by adding a spin-orbit coupling term to the 
Hamiltonian. Such magnetic moments become important, {\em e.g.}, in magnetic 
solids, where spin-orbit coupling produces phenomena such as 
magneto-crystalline an\-isotropy \cite{erikssonrev1,erikssonrev2} 
or magneto-optical effects such as dichroism \cite{ebertreview,dichroprl}. 

Finally, and most intriguingly, orbital magnetism can also occur
spontaneously in a system with pure Coulomb interactions, if the system
minimizes its energy in a current-carrying state \cite{vr2,rasoltperrot}.
The resulting currents are in principle functionals of the charge density, 
because the original formulation of the Hohenberg-Kohn
theorem applies in the absence of magnetic fields, but (S)DFT provides no
explicit prescription how to calculate the spontaneous orbital currents and
their effect on observables. This situation has changed with the advent of
nonrelativistic current-density-functional theory, developed by Vignale and 
Rasolt \cite{vr2,vr1}, which describes spontaneous currents by introducing 
in the Kohn-Sham equations a self consistent exchange-correlation ($xc$) 
vector potential ${\bf A}_{xc}$, which can be nonzero also in the absence of 
external magnetic fields and of relativistic effects.
                                                                                
In addition to possibly appearing spontaneously, ${\bf A}_{xc}$ also becomes 
nonzero as soon as currents are induced by one of the other three mechanisms. 
In this case, it constitutes a correction to the external or internal vector 
potentials, or spin-orbit terms, already present in the Hamiltonian. 

CDFT has been applied to the calculation of the effects of orbital magnetism in 
atoms \cite{pcdft1,pcdft2,tao}, quantum dots \cite{solstat4,solstat5,solstat6},
molecules \cite{handy1,handy2} and solids 
\cite{solstat1,solstat8,rasoltperrot,berlincdft2}. 
In these applications it has become clear that
the main challenge of CDFT at the present stage of its development is the
construction of reliable and computationally viable approximations to
the exchange-correlation vector potential ${\bf A}_{xc}$. The present
paper explores the possibility to develop such approximations by
drawing on analogies to independently developed phenomenological approaches
to orbital magnetism.

\section{\label{treatments}Phenomenological and first-principles treatment 
of orbital magnetism in density-functional theory}

This section collects, without proofs (which can be found in the original
literature), the key equations of both the first-principles (CDFT) and the
phenomenological (OP) approach to orbital magnetism in DFT. As our aim is to
unravel a connection between the two, we limit ourselves to providing a
brief overview, highlighting the aspects that will allow us to identify the
phenomenological approach as a well-defined approximation to the 
first-principles one.

\subsection{\label{cdft}Current-density-functional theory}

To briefly describe the formalism of nonrelativistic CDFT, we first recall the
form of the traditional Kohn-Sham (KS) equation of DFT,
\be
\left[-\frac{\hbar^2}{2m}\nabla^2 + v_s\r\right]\phi_k\r=
\eps_k \phi_k\r.
\label{dftks}
\ee
Here the effective single-particle potential, $v_s\r $, is defined as
\be
v_s\r = v\r + v_H\r + v_{xc}\r,
\ee
where $v\r$ is the external potential, $v_H\r$ the Hartree electrostatic
potential, and $v_{xc}\r$ the exchange-correlation potential, in which the
entire complexity of the many-body problem is hidden \cite{dftbook,parryang}.

By construction, the single-particle orbitals $\phi_k$ solving the eigenvalue
problem (\ref{dftks}) reproduce the density of the interacting system via
\be
n\r=\sum_k\phi_k^*\r \phi_k\r.
\ee
On the other hand, the paramagnetic current density,
\be
{\bf j}^{KS}_p\r= \frac{\hbar}{2mi}\sum_k
\left[\phi_k^*\r\nabla\phi_k\r - \phi_k\r\nabla\phi_k^*\r\right],
\label{jpks}
\ee
and the orbital magnetic moment
\be
\la \hat{L}_z\ra^{KS}
= {m \over \hbar}\int d^3r\, (\hat{\bf z} \times {\bf r}) 
\cdot {\bf j}^{KS}_p\r
\ee
following from these orbitals, are, {\em a priori}, not guaranteed
to have any relation with the true current density and magnetic moment 
of the interacting system.

The corresponding equations of CDFT have a slightly more complicated form:
\be
\left[\frac{1}{2m}\left(\frac{\hbar}{i}\nabla-\frac{q}{c}{\bf A}_s\r \right)^2
+V_s^c\r \right]\psi_k\r
=\eps_k^c \psi_k\r,
\label{cdftks}
\ee
where an upper index `$c$' denotes CDFT,
\be
V_s^c\r= v_s^c\r + \frac{q^2}{2mc^2}
\left({\bf A}\r^2 - {\bf A}_s\r^2 \right),
\label{vcdft}
\ee
\be
v_s^c\r = v\r+v_H\r+v^c_{xc}\r,
\ee
and
\be
{\bf A}_s\r={\bf A}\r + {\bf A}_{xc}\r.
\label{acdft}
\ee
Here $v^c_{xc}$ and ${\bf A}_{xc}$ are the exchange-correlation scalar and
vector potentials of CDFT, respectively \cite{vr2,vr1}.
${\bf A}_{xc}$, in particular, is a gauge invariant functional of the
densities $n\r$ and ${\bf j}_p\r$, written ${\bf A}_{xc}[n,{\bf j}_p]\r$.
By setting ${\bf A}_{xc}\equiv 0$, one recovers from CDFT the equations of 
(S)DFT in an external vector potential.
The novel feature of CDFT is that ${\bf A}_{xc}$ accounts for the
orbital degrees-of-freedom in the single-particle equations, even in
the absence of external fields, and allows calculation of orbital currents
directly from a set of Kohn-Sham-type equations.

The single-particle orbitals $\psi_k$ solving the more complicated 
eigenvalue problem (\ref{cdftks}) reproduce by construction both the density
\be
n\r=\sum_k\psi_k^*\r \psi_k\r,
\ee
{\em and} the paramagnetic current density
\be
{\bf j}_p\r= \frac{\hbar}{2mi}\sum_k
\left[\psi_k^*\r\nabla\psi_k\r-(\nabla\psi_k^*\r)\psi_k\r\right]
\label{jcdft}
\ee
of the interacting many-body system. The correct orbital magnetic moment can
then be obtained from
\be
\la \hat{L}_z\ra 
= {m \over \hbar}\int d^3r\, (\hat{\bf z} \times {\bf r}) \cdot {\bf j}_p\r.
\label{lzfuncjp}
\ee

The paramagnetic current alone is not gauge invariant, but the gauge invariant 
orbital current, ${\bf j}_{\rm orb}\r$, is simply obtained from
\be
{\bf j}_{\rm orb}\r = {\bf j}_p\r -\frac{q}{mc}n\r{\bf A}\r.
\label{fullj}
\ee
An important property of the effective potential of DFT, $v_s\r$, which is
{\em not} shared by the CDFT potentials $V_s^c\r$ and ${\bf A}_s\r$
\cite{nonunprb}, is its {\em uniqueness}: for any given ground-state density 
$n\r$ there is, up to an irrelevant additive constant, at most {\em one} such 
local multiplicative potential \cite{dftbook}.

Any CDFT calculation requires an approximation for the current
dependence of the $E_{xc}$ functional. For the homogeneous
three-dimensional electron liquid in strong uniform magnetic fields,
the exchange energy is known exactly \cite{danzglasser}, and the
correlation energy has been calculated within the random-phase
approximation \cite{cdftrpa} and the self-consistent local-field
corrected Singwi-Tosi-Land-Sjolander scheme \cite{cdftstls}. In weak
fields, where linear-response theory applies, the exchange-correlation
energy can be expressed in terms of the magnetic susceptibility, for
which many-body calculations are available from \cite{vrg}.  The
exchange-correlation energy of two-dimensional electron liquids in
uniform magnetic fields has been much studied in the context of the
fractional Quantum Hall effect in quasi-two-dimensional semiconductor
heterostructures \cite{rossler}.

\subsection{\label{brooks}Orbital-polarization terms}

In an independent take on the problem of orbital magnetism, originating
with Brooks and collaborators \cite{brookspaper,eriksson}, phenomenological
orbital-polarization terms are introduced in the formal framework of
SDFT. This effort is motivated by
observing that SDFT does account for Hund's first rule (dealing with spin
angular momentum) and can be extended to account for Hund's third rule
(due to spin-orbit coupling), but does not contain an obvious ingredient 
corresponding to Hund's second rule (dealing with orbital angular momentum).

In the absence of external magnetic fields, the exact exchange-correlation 
functional of SDFT doubtlessly would predict the correct orbital magnetic 
moments and orbital currents, as {\em functionals} of the spin and charge 
densities, but these functionals are not known. In practice, such quantities 
are therefore often calculated directly from the {\em orbitals} of SDFT. 
We note that there are two separate issues here: One is the use of SDFT 
orbitals to calculate orbital magnetic moments, although these orbitals are 
constructed to reproduce only the charge and spin densities, not the orbital 
currents. The other is the use of an approximate SDFT functional (the LSDA) 
in these calculations.

In practice, the resulting orbital magnetic moments strongly underestimate 
experimental values for transition metals \cite{ebert2}, and 
make wrong predictions for, {\em e.g.}, the volume collapse in lanthanides 
\cite{eriksson} and the band gap of transition-metal oxides 
\cite{norman1,norman2}. This behaviour was interpreted as a consequence of 
the inapplicability of Hund's second rule to the electron gas, on which the 
LSDA is based. To account for Hund's second rule, phenomenological 
orbital angular momentum dependent corrections to the LSDA were proposed by 
drawing on analogies with angular momentum dependent terms in the multiplet 
splitting of atoms \cite{brookspaper,eriksson}. Significant improvement of 
magnetic moments and related quantities is obtained from this so-called 
orbital-polarization approach \cite{brookspaper,eriksson,erikssonrev1,erikssonrev2,ebert2,norman1,angela,sonia,solstat1,eschrig,ebert1}.

In Sec.~\ref{connections}, we explain that this empirical success can be
understood if the OP term is not considered a better SDFT functional than 
the LSDA, but rather an approximate CDFT functional. As such, it produces
orbitals that can reproduce charge, spin and current densities, and thus
also orbital magnetic moments. 

The original proposal for systems with open shells of $4f$ electrons was to 
add the OP term (today also known as Brooks term) 
\begin{equation}
\Delta E^{B} = -\frac{1}{2}E^{3} L^2
\label{brooks-energia}
\end{equation}
to the total-energy functional of SDFT. Here $E^{3}$ is a Racah parameter
defined in terms of Slater integrals $F_k$ as 
$E^{3}=(5F_2+6F_4-91F_6)/3$ \cite{racah,griffith}, and
$L$ is the total orbital angular momentum of the open-shell ion defined as
\cite{eriksson,norman1,erikssonrev1}
\be
L:= \sum_i \la \hat{l}_z \ra 
= \la \hat{L}_z \ra
= \sum_{nlm_{l}\sigma}{\gamma_{nlm_{l}\sigma}m_{l}} 
\label{lcalc}
\ee 
in terms of a sum over occupation numbers $\gamma_{nlm_l\sigma}$ of 
single-particle orbitals. 
Expressions (\ref{brooks-energia}) and (\ref{lcalc}) are motivated by noting 
the presence of similar terms in the vector model of the multiplet splitting 
of atoms with open $f$ shells \cite{eriksson,norman1,norman3,racah,griffith}, 
but their inclusion in an SDFT calculation is entirely {\em ad hoc}. 
Nevertheless,
by adopting Eq.~(\ref{brooks-energia}) as a starting point, simple and
empirically successful correction terms to the single-particle equations of
SDFT are obtained \cite{brookspaper,eriksson,erikssonrev1,erikssonrev2,ebert2,norman1,angela,sonia,solstat1,eschrig,ebert1}.

The original OP idea was subsequently extended to $d$-electrons, for which
Norman proposed \cite{norman1}
\begin{equation}
\Delta E^{Nd'} = -\frac{1}{2} B L^2,
\label{norman-energia}
\end{equation}
where $B$ is the Racah parameter $B=F_2-5F_4$ \cite{racah,griffith}.

Further refinements are based on a more complete treatment of atomic multiplet 
splitting \cite{norman3,shick} and crystal-field effects \cite{norman2}.
For $d$-electrons, Norman \cite{norman3} proposed the expression
\be
\Delta E^{Nd} = 2.25n_{d}(5-n_{d})B - 1.5 L(L+1)B,
\label{nd2}
\ee
where $n_d$ is the occupation number of the $3d$ orbitals, and for 
$f$-electrons,
\be
\Delta E^{Nf} = (-2n_{f}^2 + 14n_{f})E^{3} - L(L+1 )E^{3},
\label{nf}
\ee
where $n_f$ is the occupation number of the $4f$ orbitals.
Shick e Gubanov \cite{shick} refined Norman's argument for $f$-electrons, and 
obtained 
\be
\Delta E^{SG} = -{3\over 2}\left[L(L +1 ) - 6 g(G_{2})\right] E^{3}, 
\label{sgf}
\ee
where $g(G_{2}) = [6n_{f}(7-n_{f}) +41n_{f}^{2}
(7-n_{f})^{2} - 2n_{f}^{3}(7-n_{f})^{3}]/(2^23^35)$.
This expression by construction recovers the correct splitting of the
highest-spin multiplet of the isolated ion \cite{norman3,shick}.

Although expressions (\ref{nd2}), (\ref{nf}), and (\ref{sgf}) give a better 
description of atomic multiplets than Eqs.~(\ref{brooks-energia}) and 
(\ref{norman-energia}), applications to solids containing transition-metal 
and rare-earth atoms \cite{eriksson,erikssonrev1,erikssonrev2,ebert2,norman1,angela,sonia,solstat1,ebert1} mostly employed the simpler expressions 
(\ref{brooks-energia}) and (\ref{norman-energia}),
which already lead to substantial improvement of orbital magnetic moments 
and related quantities, as compared to uncorrected LSDA. 
In Sec.~\ref{connections}, we therefore mainly focus on the simple expressions 
(\ref{brooks-energia}) and (\ref{norman-energia}).

We note in passing that apparently no OP term has been proposed for $p$
electrons, although the physical motivation and the mathematical argument
leading to the OP terms for $d$ and $f$ electrons would apply to them,
too. From an analysis of the multiplet splitting of the $p$ configurations
\cite{griffith}, we find that the $p$-electron OP term should have
the form
\be
\Delta E^p=\frac{3}{2}\left[n_p(6-n_p)-L(L+1)-4S(S+1)\right] F_2,
\label{pop1}
\ee
which is obtained by subtracting from the energy of the terms arising 
from the $p$-shell their spherical average. 
This expression for $p$ electrons is valid for all $S$, in contrast to
expressions (\ref{nd2}) for $d$ \cite{norman3} and (\ref{nf}) and (\ref{sgf})
for $f$ \cite{norman3,shick} electrons, which hold only for the
highest spin multiplet, where $S=n_d/2$ and $S=n_f/2$, respectively.
In analogy to Eqs.~(\ref{brooks-energia}) and (\ref{norman-energia}), a 
simpler version, expected to simulate the effect of (\ref{pop1}), is
\be
\Delta E^{p'} = - {1\over 2} F_2 L^2.
\label{pop2}
\ee

Finally, we remark that $d$ and $f$-electron OP terms of similar form to 
(\ref{nd2}), (\ref{nf})
and (\ref{sgf}) have been obtained also within relativistic DFT
\cite{eschrig}. In that work, as well as in Refs.~\cite{ebert1,ebert2},
it was pointed out that there should be a connection between the 
various phenomenological orbital-polarization corrections and the 
first-principles formalism of (relativistic or nonrelativistic) CDFT. 
In the next section we demonstrate this connection.

\section{\label{connections}Connection between the phenomenological and the 
first-principles approach}

In this section we establish a connection between the OP terms and CDFT.
This connection is obtained in two different ways, once by analysing 
single-particle Hamiltonians (Sec.~\ref{eps}), and once by
considering total energies (Sec.~\ref{Es}). The method of analysis is 
different in both cases, but the final results are the same.

\subsection{\label{eps}Single-particle energies}

In the absence of external magnetic fields, the Kohn-Sham single-particle
Hamiltonians of CDFT and DFT take the form
\begin{equation}
\hat{H}^{c}=-\frac{\hbar^{2}}{2m}\nabla^{2}-\frac{i\hbar e}{2mc}\left[{\bf A}_{xc}({\bf r})\cdot \nabla + \nabla \cdot {\bf A}_{xc}({\bf r})\right]
+ v({\bf r})+v_{H}({\bf r})+v_{xc}^{c}({\bf r})
\label{cdft-brooks}
\end{equation}
and
\begin{equation}
\hat{H}=-\frac{\hbar^{2}}{2m}\nabla^{2}+v({\bf r})+v_{H}({\bf r})+v_{xc}({\bf r}),
\label{sdft-brooks}
\end{equation}
respectively. There difference is thus given by the operator
\begin{equation}
\Delta \hat{H} = \hat{H}^{c}-\hat{H}=-\frac{i\hbar e}{2mc}\left[{\bf A}_{xc}({\bf r})\cdot \nabla + \nabla \cdot {\bf A}_{xc}({\bf r})\right] +
\Delta v_{xc}\r,
\end{equation}
where $\Delta v_{xc}\r = v_{xc}^{c}({\bf r}) - v_{xc}({\bf r})$. A perturbative
treatment of the first term of this difference was suggested in 
Ref.~\cite{jpert} and applied mainly to $p$ electrons in \cite{pcdft1,pcdft2}. 
In the context of magnetism, however, our main interest is in $d$ and $f$ 
electrons, for which the difference between both Hamiltonians may be too large 
to justify a low-order perturbation treatment. Moreover, the functional used
in \cite{pcdft1,pcdft2}, as well as in many other applications of CDFT
\cite{handy1,handy2,solstat1}, is based on the electron liquid,
which at stronger magnetic fields displays quantum oscillations 
\cite{cdftrpa,cdftstls}, which are incorrect for finite atomic and molecular
systems.

Independently of the size of the various terms, the identity
$\nabla\cdot\left[n({\bf r}){\bf A}_{xc}({\bf r})\right]=0$ 
always holds \cite{vr2,vr1}, and can be used to write $\Delta \hat{H}$
in spherical polar coordinates ($r, \theta, \varphi$) as
\bea
\Delta \hat{H} = \Delta v_{xc}\r - \nonumber \\
\frac{i \hbar e}{2mc}\left[2\left(A_{xc}^{r}({\bf r})\frac{\partial}{\partial r} 
+ \frac{1}{r}A_{xc}^{\theta}({\bf r})\frac{\partial}{\partial \theta} + 
\frac{1}{r\sin{\theta}}A_{xc}^{\varphi}({\bf r})\frac{\partial}{\partial \varphi}\right)
-\frac{\nabla n({\bf r})}{n({\bf r})}\cdot {\bf A}_{xc}({\bf r})\right].
\label{deltaH}
\eea
This is the complete set of (nonrelativistic) corrections to DFT 
predicted by CDFT in the absence of external fields. 

We can now compare this to the various {\em ad hoc} corrections listed in 
Sec.~\ref{brooks}. According to Janak's theorem \cite{janak}, the DFT 
Kohn-Sham eigenvalue is obtained from 
the total energy by differentiating with respect to the occupation number,
\be
\eps_{nlm_{l}\sigma} = \frac{\partial E}{\partial \gamma_{nlm_{l}\sigma}}.
\ee
If this relation, which also holds in SDFT and CDFT, is applied to the SDFT
functional augmented by the orbital polarization term (\ref{brooks-energia}),
and the Racah parameter is kept fixed during the differentiation,
it follows that the {\em ad hoc} OP correction to the total energy corresponds
to a correction to the single-particle energies of the form
\begin{equation}
\Delta \epsilon^{B} = -E^{3} L m_{l},
\label{brooks-autovalor}
\end{equation}
which can in turn can be interpreted as a result of the operator
\begin{equation}
\Delta \hat{H}^{B} = -E^{3} L\, \hat{l}_{z},
\label{brooks-operador}
\end{equation}
added to the single-particle equation of SDFT. Similarly, 
\begin{equation}
\Delta \hat{H}^{Nd'} = -BL\,\hat{l}_{z},
\label{norman-operador0}
\end{equation}
\begin{equation}
\Delta \hat{H}^{p'} = -F_2L\,\hat{l}_{z},
\label{p-operador0}
\end{equation}
\begin{equation}
\Delta \hat{H}^{Nd} = 2.25\left(5-2n_{d}\right)B -
3B\left(L +\frac{1}{2}\right)\hat{l}_{z},
\label{normand-operador}
\end{equation}
\begin{equation}
\Delta\hat{H}^{Nf}=
\left(-4n_{f}+14\right)E^{3} - 2E^{3}\left(L +\frac{1}{2}\right)\hat{l}_{z},
\label{normanf-operador}
\end{equation}
and
\begin{eqnarray}
\Delta \hat{H}^{SG} =
 -3E^{3}\left(L +\frac{1}{2}\right)\hat{l}_{z} + \frac{9}{2^{2} 3^{3} 5}
\nonumber \\
\left[\left(7-n_{f}\right)\left(6-82n_{f}^{2}\right) +
\left(7-n_{f}\right)^{2}\left(82n_{f}+6n_{f}^{3}\right) 
-6n_{f}^{2}\left(7-n_{f}\right)^{3}-6n_{f}\right]E^{3}. 
\label{shick-operador}
\end{eqnarray}

We note that if the Racah term is not treated as a number, but as a matrix
element involving single-particle radial orbitals \cite{racah,griffith},
additional terms appear on the right-hand side of these equations. 
(We return to this issue at
the end of Sec.~\ref{Es}, from a slightly different point of view.)

Since $\hat{l}_{z} = -i \partial / \partial \varphi$, 
the single-particle operators resulting from the OP terms, Eqs.~(\ref{brooks-operador}), (\ref{norman-operador0}) and (\ref{p-operador0}), 
can be cast in the form of Eq.~(\ref{deltaH}), with
\begin{equation}
{\bf A}_{xc}^{OP}({\bf r}) = 
-\frac{mc}{\hbar e} R_{l}\, r L \sin{\theta}\, \hat{\varphi},
\label{axc-op}
\end{equation}
along with
\begin{equation}
v_{xc}^{c,OP}({\bf r}) = v_{xc}({\bf r}),
\label{vxc-op}
\end{equation}
and
$A_{xc}^{OP,r}({\bf r}) = A_{xc}^{OP,\theta}({\bf r}) = 
\nabla n({\bf r}) \cdot {\bf A}^{OP}_{xc}({\bf r}) = 0$. In these equations
$\hat{\varphi}$ is the unit vector in the azimuthal direction, and
$R_l=E^3$,$B$,$F_2$, for $f$, $d$, and $p$ electrons, respectively.

This way of writing the OP correction shows that it enters the KS
Hamiltonian in the same way as the $xc$ vector potential of CDFT.
To complete the proof that it really is such a potential, 
we need to write it as a functional of the current density, 
${\bf A}_{xc}^{OP}[{\bf j}_p\r]$, which is easily accomplished by 
means of Eqs.~(\ref{lzfuncjp}) and (\ref{lcalc}).

We note that $v_{xc}^{c,OP}({\bf r}) = v_{xc}({\bf r})$ implies
$\Delta v^{OP}_{xc}\r =0$, which may be interpreted as a weak 
coupling approximation, in which the self-consistent effect of currents on 
the electric potential is considered small.
The neglect of the radial and polar components of ${\bf A}_{xc}\r$ is
correct for systems with spherical symmetry, where $\nabla n({\bf r}) \cdot
{\bf A}_{xc}$ also vanishes. The orbital-polarization terms can thus 
be identified as weak-coupling, spherically symmetric approximations to the 
self-consistent $xc$ vector potential of CDFT. 

This conclusion was anticipated in Refs.~\cite{ebert2,ebert1}, where it was,
however, argued that the physics described by the Brooks term was distinct
from that of CDFT. Our present point of view is different: CDFT is a 
general framework for describing {\em all} nonrelativistic effects of
orbital magnetism, and to the extent that the orbital polarization terms
are justifiable, they must be specific approximations to the general 
framework of CDFT.

The other orbital-polarization terms can also be identified as approximations 
to ${\bf A}_{xc}$. Eqs.~(\ref{normand-operador}) and (\ref{normanf-operador}) lead to
\begin{equation}
{\bf A}_{xc}^{Nd}({\bf r}) = -3 \frac{mc}{\hbar e}B\left(L +\frac{1}{2}\right) r
\sin{\theta} \ \hat{\varphi}
\label{axc-normand}
\end{equation}
and 
\begin{equation}
v_{xc}^{c,Nd}({\bf r}) = v_{xc}({\bf r}) + 2.25\left(5-2n_{d}\right)B
\label{vxc-normand}
\end{equation}
for $d$ electrons, and
\begin{equation}
{\bf A}_{xc}^{Nf}({\bf r}) = -2 \frac{mc}{\hbar e}E^{3}\left(L +\frac{1}{2}\right) r
\sin{\theta} \ \hat{\varphi}
\label{axc-normanf}
\end{equation}
and
\begin{equation}
v_{xc}^{c,Nf}({\bf r}) = v_{xc}({\bf r}) + \left(-4n_{f}+14\right)E^{3}
\label{vxc-normanf}
\end{equation}
for $f$ electrons. Interestingly, the scalar $xc$ potentials are also
modified by these proposals. However, the modification is a constant shift,
independent of position, whereas $\Delta v_{xc}\r=v_{xc}^c({\bf r})-
v_{xc}({\bf r})$ should, generically, depend on ${\bf r}$.

Finally, the proposal of Shick e Gubanov implies
\begin{equation}
{\bf A}_{xc}^{SG}({\bf r}) 
= -3 \frac{mc}{\hbar e}E^{3}
\left(L +\frac{1}{2}\right) r \sin{\theta} \ \hat{\varphi}
\label{axc-shick}
\end{equation}
and
\bea
v_{xc}^{c,SG}({\bf r}) = v_{xc}({\bf r}) + \frac{E^{3}}{30}\left[21+2003n_{f}-1890n_{f}^{2}+670n_{f}^{3}-105n_{f}^{4}+6n_{f}^{5}\right]
\label{vxc-shick}.
\eea

All these expressions for ${\bf A}_{xc}\r$ and $v_{xc}^c\r$ can be implemented
self-consistently, by recalculating the Racah parameters and the 
occupation numbers at every iteration, although such a self-consistent 
recalculation is not consistent with treating these parameters as numbers when
deducing the eigenvalue correction from the total-energy correction.

\subsection{\label{Es}Total energies}

The connection between the CDFT $xc$ potentials and the orbital 
polarization corrections, summarized in Eqs.~(\ref{axc-op}) and
(\ref{vxc-op}) was obtained by comparing operators in the 
single-particle equations of CDFT and of the SDFT + OP approach. 
The operator in the SDFT + OP equations was obtained by means of Janak's 
theorem, from partial differentiation of the total energy with respect 
to occupation numbers.

Alternatively, one can follow the established DFT way to obtain potentials 
directly from total energies, by variational differentiation with respect to 
densities, according to \cite{vr1, vr2}
\be
v_{xc}^c({\bf r}) = 
\frac{\delta E_{xc}^c[n,  {\bf j}_{p}]}{\delta n({\bf r})}
\label{vxc-vignale}
\ee
and 
\be
\frac{e}{c}{\bf A}_{xc}({\bf r}) = 
\frac{\delta E_{xc}^{c}[n,  {\bf j}_{p}]}{\delta {\bf j}_{p}({\bf r})}.
\label{axc-vignale}
\ee
Application of these definitions to the DFT total energy expression, 
augmented by the OP terms (\ref{brooks-energia}), (\ref{norman-energia}) or (\ref{pop2}), leads to
\bea
v_{xc}^{c,OP}({\bf r}) &=& \frac{\delta E_{xc}[n]}{\delta n({\bf r})} -
\frac{1}{2}\frac{\delta}{\delta n({\bf r})}\left[R_{l}\langle \hat{L}_{z} \rangle^{2} \right]  
\nonumber \\
& = & v_{xc}({\bf r}) -
\frac{1}{2}\frac{\delta}{\delta n({\bf r})}\left[R_{l}\langle\hat{L}_{z} \rangle^{2} \right]
\eea
and
\bea
\frac{e}{c}{\bf A}_{xc}^{OP}({\bf r}) & = & \frac{\delta E_{xc}[n]}{\delta {\bf j}_{p}({\bf r})} -
\frac{1}{2}\frac{\delta}{\delta {\bf j}_{p}({\bf r})}\left[R_{l}\langle \hat{L}_{z} \rangle^{2} \right]
\nonumber \\
&=& 
0 -R_{l} \langle \hat{L}_{z} \rangle \frac{\delta}{\delta {\bf j}_{p}({\bf r})}\langle \hat{L}_{z} \rangle -
\frac{1}{2}\langle \hat{L}_{z} \rangle^{2} \frac{\delta R_{l}}{\delta {\bf j}_{p}({\bf r})}.
\eea
These equations can be simplified by using Eq.~(\ref{lzfuncjp}) to calculate
$\delta \langle \hat{L}_{z} \rangle / \delta {\bf j}_p\r$. Since within CDFT 
$n\r$ and ${\bf j}_p\r$ are independent variables, we arrive at
\begin{equation}
v_{xc}^{c,OP}({\bf r}) = v_{xc}({\bf r}) - \frac{1}{2}\langle \hat{L}_{z} \rangle^{2} \frac{\delta
R_{l}}{\delta n({\bf r})}
\label{vxc-brooks-vignale}
\end{equation}
and
\begin{equation}
{\bf A}_{xc}^{OP}({\bf r}) = -\frac{mc}{\hbar e}R_{l} \ r \sin{\theta} \ \langle \hat{L}_{z} \rangle \
\hat{\varphi} - \frac{c}{2e} \langle \hat{L}_{z} \rangle^{2} \frac{\delta R_{l}}{\delta
{\bf j}_{p}({\bf r})}.
\label{axc-brooks-vignale}
\end{equation}
If we again treat the Racah parameters as numbers, and not as self-consistent 
functionals of the densities, we finally obtain
\begin{equation}
v_{xc}^{c,OP}({\bf r}) = v_{xc}({\bf r}) 
\label{vxc-brooks2}
\end{equation}
and
\begin{equation}
{\bf A}_{xc}^{OP}({\bf r}) = 
-\frac{mc}{\hbar e}R_{l} \ r \sin{\theta} \langle \hat{L}_{z} \rangle \hat{\varphi}.
\label{axc-brooks2}
\end{equation}

These are the same relations obtained in the preceding section from
Janak's theorem and partial differentiation with respect to occupation
numbers. Just as there, the Racah parameters must be treated as fixed 
numbers in order to arrive at the standard form of the OP term.
We note that if the Racah coefficients are differentiated consistently
also in the derivation via Janak's theorem, one obtains, by application of
the chain rule in the form
\be
\frac{\partial R_l}{\partial \gamma} =
\int d^3r\, \left[\frac{\delta R_l}{\delta n\r}\frac{\partial n\r}{\partial \gamma}
+ \frac{\delta R_l}{\delta {\bf j}_p\r}\frac{\partial {\bf j}_p\r}{\partial \gamma}
\right],
\ee
the same extra
terms appearing on the right-hand side of Eqs.~(\ref{vxc-brooks-vignale})
and (\ref{axc-brooks-vignale}). To the best of our knowledge, consequences 
of these additional $\langle L_z\rangle^2$ term have never been systematically 
explored.

\section{\label{conclusions}Conclusions}

The $f$-electron orbital polarization term proposed by Brooks at al.
\cite{brookspaper,eriksson}, its generalization to $d$ electrons 
\cite{norman1}, and their refinements and generalizations
\cite{norman2,norman3,shick}, have been identified as specific 
approximations to ${\bf A}_{xc}\r$ and $v_{xc}^c\r$ of CDFT. This
identification provides formal justification for their introduction into
the Kohn-Sham Hamiltonian of (S)DFT: by using an orbital polarization
correction one actually simulates certain aspects of CDFT within the
computational framework of (S)DFT. 

The existence of such a connection implies that the OP terms should not
be interpreted just as corrections to the LSDA of SDFT, although this may have
been their original motivation, but also (and perhaps principally!) as a step 
from SDFT to CDFT. Hence, from the present point of view, the OP terms address 
both issues facing the calculation of orbital moments in SDFT that were
mentioned at the beginning of Sec.~\ref{brooks}: correction of 
electron-gas-based $xc$ functionals for Hund's second rule, and substitution
of the SDFT orbitals by CDFT orbitals that reproduce the orbital currents.
We stress that this also holds for the $p$-electron OP terms (\ref{pop1})
and (\ref{pop2}), proposed in this work.

Explicit expressions for ${\bf A}_{xc}$ and $v_{xc}^c\r$ that can be used 
in CDFT are scarce, and mostly based on many-body calculations of
the energy of the uniform electron gas in certain ranges of external
magnetic fields \cite{vr2,vrg,danzglasser,cdftrpa,cdftstls,rossler}.
Explicit, albeit approximate, expressions for ${\bf A}_{xc}$ and $v_{xc}^c\r$
that by construction go beyond local-density approximations may be useful
in applications to finite systems, and constitute starting points for the
development of more refined functionals.

As an example, by combining (\ref{axc-brooks2}) with (\ref{lzfuncjp}) we
obtain the explicit current-density functional
\be
{\bf A}_{xc}^{OP}[{\bf j}_p]({\bf r}) 
= -\frac{m^{2}c}{\hbar^{2} e} R_l \, r \sin{\theta}\left[\int d^3r \, r \sin{\theta} j_p^{\varphi}\r \right] \hat{\varphi}.
\label{axcfunctional}
\ee

The empirical success of the OP concept in solid-state physics suggests 
that this may be a useful and reliable expression for ${\bf A}_{xc}$ of 
molecular and solid systems containing open-shell atoms. As this expression is 
nonperturbative, and not based on the electron liquid, it is expected to 
be more appropriate for systems containing open-shell atoms than the 
linear-response LDA \cite{vr2,handy1,handy2,pcdft1,pcdft2,solstat1}.
This expression also does not suffer from quantum oscillations that stem
from the electron liquid but are incorrect for the atomic and molecular
systems \cite{cdftrpa,cdftstls}. It may therefore constitute a useful starting 
point also for applications of CDFT to molecular systems.\\

{\bf Acknowledgments} 
This work was supported by FAPESP and CNPq. KC thanks Angela Klautau and
Helmut Eschrig for useful conversations about orbital polarization terms.

\end{document}